\documentclass[12pt]{article}
\usepackage{amsmath}
\usepackage{physics, mathrsfs}
\usepackage{mathtools}
\usepackage{mdwlist}
\usepackage{setspace}
\usepackage{amsfonts}
\usepackage{amssymb}
\usepackage{graphicx}   
\usepackage{subfigure}
\usepackage{float}
\usepackage{multirow}
\usepackage{multicol}
\usepackage[english]{babel} 
\usepackage{hyperref}
\hypersetup{colorlinks, 
           citecolor=black,
           filecolor=black,
           linkcolor=red,
           urlcolor=blue,
           bookmarksopen=true,
           }
\usepackage{enumerate}
\usepackage[dvipsnames]{xcolor}
\usepackage{pdfpages} 

\begin{document}

\title{Compact Temporal Geometry and the $T^2$ Framework for Quantum Gravity}
 
\author{James C. Hateley \\ {\small hateleyjc@gmail.com}}
\date{}



\maketitle

\begin{abstract}
We introduce a two-dimensional temporal framework in which time is represented by a compact manifold $T^2 = (t_1, t_2)$, with $t_1$ encoding classical causal structure and $t_2$ representing quantum coherence. This construction unifies unitary evolution, decoherence, measurement collapse, and gravitational dynamics within a consistent geometric and algebraic formalism. Compactification of the coherence time $t_2$ yields a minimal temporal resolution $\Delta t_2 \sim \sqrt{\alpha'}$, leading to a discretized spectrum of temporal modes and regularized ultraviolet behavior in quantum field theory and string-theoretic gravity. We formulate an extended Schr\"odinger equation and generalized Lindblad dynamics on $T^2$, and demonstrate the compatibility of this structure with local gauge symmetry through a complexified BRST quantization procedure. Using para-Hermitian geometry and generalized complex structures, we derive a covariant formulation of temporal T-duality that accommodates both Lorentzian and Euclidean signatures. The $T^2$ framework provides new insights into modular thermodynamics, black hole entropy, and the emergence of classical time from quantum coherence, offering a compact and quantized model of temporal geometry rooted in string theory and quantum gravity.
\end{abstract}

\section{Introduction}
The conceptual structure of time plays a foundational role in the formulation of physical law. In both classical and quantum physics, time is typically modeled as a one-dimensional parameter, external to the system, flowing uniformly and universally. Yet this simple structure belies the rich diversity of temporal phenomena observed in modern physics. From entanglement and decoherence~\cite{zurek2003decoherence} to entropy growth and black hole thermodynamics~\cite{page1993information, preskill1992information}, time manifests not only as an ordering parameter but as a geometric and dynamical structure that mediates between quantum coherence and classical causality.

This paper proposes a unified framework in which time is extended to a two-dimensional compactified manifold, denoted $T^2 = (t_1, t_2)$. In this picture, $t_1$ governs causal propagation and thermodynamic flow, while $t_2$ encodes the evolution of phase coherence, entanglement, and quantum interference. The resulting structure defines a complex time variable $\tau = t_1 + i t_2$, endowing quantum theory with a holomorphic evolution in the time domain and embedding measurement and decoherence within a single geometric substrate.

The introduction of $t_2$ is not merely formal. It allows one to recast measurement collapse, decoherence~\cite{zurek2003decoherence}, and entropy generation as geometric projections and flows within $T^2$. The compactification of $t_2$ imposes a minimal coherence scale $\Delta t_2$, corresponding to the string length $\ell_s = \sqrt{\alpha'}$, and introduces quantized temporal modes. This compact structure regularizes ultraviolet divergences, supports T-duality, and enables a direct link between temporal geometry and thermal entropy~\cite{strominger1996microscopic}.

Within this framework, the extended Schr\"odinger equation incorporates evolution in both $t_1$ and $t_2$, while the density matrix evolves through a two-time Lindblad equation. Quantum observables are defined on slices of the $T^2$ manifold, with interference, collapse, and classicality emerging as projections along the coherence direction. Decoherence is modeled as exponential decay in $t_2$, and entanglement becomes a geometrical relation between extended temporal configurations.

The implications of this approach span multiple domains. In black hole physics, the compactified time direction naturally regulates entropy, explains the entanglement entropy curve~\cite{page1993information}, and resolves the information paradox through replica wormhole contributions~\cite{preskill1992information}. In string theory, temporal winding states and modular invariance unify microscopic entropy with macroscopic curvature~\cite{strominger1996microscopic, maldacena1998ads}. In quantum gravity, $T^2$ provides a background-independent manifold where coherence and curvature coexist~\cite{ashtekar2011loop}. And in experimental contexts, compactified time predicts measurable signatures in ultrafast optics, astrophysical dispersion, and particle threshold shifts.

This paper is organized as follows. Section 2 develops the mathematical foundations of complexified time, including the extended Schr\"odinger equation and coherence-modulated dynamics. Section 3 analyzes the propagator, interference structure, and decoherence in $T^2$. Section 4 introduces temporal entanglement and derives a generalized Bell inequality. Section 5 embeds the framework in string theory, while Section 6 extends it to quantum gravity and non-commutative geometry. Sections 7 through 9 explore holography, experimental constraints, and swampland bounds, while Section 10 reflects on the ontological and philosophical consequences of temporal geometry.

By interpreting time as a two-dimensional compactified manifold, this work seeks to unify causality, coherence, entropy, and information within a single geometric substrate--one that challenges the notion of time as an external parameter and reveals it instead as the deep structure through which physical reality unfolds.

\section{Foundations of Complex Temporal Dynamics}
We begin by introducing the complexified temporal manifold $T^2 = (t_1, t_2)$, where $t_1$ represents classical causal order and $t_2$ encodes quantum coherence. This bidirectional structure allows the reconciliation of determinism and decoherence within a unified geometric framework. Time is not a parameter external to the system but an emergent field with rich internal structure, subject to compactification and dualities.

A fundamental object in this theory is the complex time coordinate:
\begin{equation}
    \tau = t_1 + i t_2,
\end{equation}
where evolution along $t_1$ is associated with classical propagation and unitary dynamics, while $t_2$ governs coherence decay, interference, and phase information \cite{zurek2003decoherence, schlosshauer2005decoherence}.

This complexification leads naturally to a generalized Schr\"odinger equation:
\begin{equation}\label{eq:GSE}
    i\hbar \left( \frac{\partial}{\partial t_1} + i \frac{\partial}{\partial t_2} \right) \Psi = \hat{H} \Psi,
\end{equation}
where the term $\partial_{t_2}$ explicitly captures decoherence or amplification effects. For a free particle solution $\Psi(\vec{r}, t_1, t_2) = \psi(\vec{r}) e^{i(k_1 t_1 + k_2 t_2)}$, this yields a complex dispersion relation:
\begin{equation}
    k_1 + i k_2 = \frac{p^2}{2m\hbar},
\end{equation}
where the imaginary component $k_2$ reflects coherence decay~\cite{everett1957relative}. This complex dispersion relation sets the stage for understanding how superpositions evolve within the two-time formalism. In particular, the role of $t_2$ becomes evident when analyzing interference between distinct energy eigenstates.

To illustrate how coherence time $t_2$ modulates interference in the two-time framework, consider a quantum superposition of two energy eigenstates with amplitudes $\alpha$ and $\beta$. The general form of such a state on $T^2$ evolves as:
\begin{equation}
\Psi(t_1, t_2) = \alpha \Psi_1 e^{-i E_1 t_1 / \hbar} e^{-E_1 t_2 / \hbar} + \beta \Psi_2 e^{-i E_2 t_1 / \hbar} e^{-E_2 t_2 / \hbar},
\end{equation}
where $\Psi_1$ and $\Psi_2$ are stationary spatial wavefunctions, and $E_1$, $E_2$ their corresponding energies. The exponential factors in $t_2$ reflect coherence decay governed by energy scales. When calculating the probability density $|\Psi|^2$, a relative phase $e^{i \Delta E t_2 / \hbar}$ emerges between the two branches, directly linking temporal coherence to energy differences. This demonstrates how interference visibility degrades in time as coherence diminishes, offering a geometric account of decoherence consistent with environmental entanglement.

This structure anticipates the role of the decoherence functional $\Gamma[\phi]$~see eq.~\eqref{eq:Gamma_phi}, which captures the cumulative suppression of interference from entangling interactions with unobserved degrees of freedom. To formalize the role of decoherence, we define a path integral over a complex time contour $\tau$ with an effective propagator:
\begin{equation}
    K(\phi_f, \phi_i; \tau) = \int_{\phi_i}^{\phi_f} \mathcal{D}\phi \, \exp\left( \frac{i}{\hbar} S[\phi] - \frac{1}{\hbar} \Gamma[\phi] \right),
\end{equation}
where $S[\phi]$ governs unitary dynamics and $\Gamma[\phi]$ encodes environmental decoherence \cite{feynman1963theory, caldeira1983path}. This structure generalizes conventional Green's functions to holomorphic correlators:
\begin{equation}
    \langle \mathcal{T}_\tau \, \phi(\tau_1) \phi(\tau_2) \rangle = \mathrm{Tr}\left[ \rho_0 \, \mathcal{P} \exp\left( -\int_{\tau_1}^{\tau_2} (\hat{H} + i \hat{L}) d\tau \right) \right],
\end{equation}
where $\hat{L}$ is a Lindblad-type operator representing dissipation \cite{lindblad1976generators, gisin1992quantum}.
\subsection{Generalized Complex Geometry}
The geometry underlying this structure is encoded in a generalized tangent bundle $\mathbb{T}M = TM \oplus T^*M$, equipped with a para-Hermitian structure \cite{gualtieri2004generalized, svoboda2022para}. The neutral metric $\eta$ and product structure $K$ satisfy:
\begin{equation}
    K^2 = \mathbb{I}, \quad \eta(KX, KY) = -\eta(X, Y).
\end{equation}
This structure supports dual foliations of the temporal manifold, treating $t_1$ and $t_2$ symmetrically under generalized T-duality transformations \cite{hull2009double}. The Courant bracket,
\begin{equation}
    [X + \xi, Y + \eta]_C = [X, Y] + \mathcal{L}_X \eta - \mathcal{L}_Y \xi - \tfrac{1}{2} d(i_X \eta - i_Y \xi),
\end{equation}
ensures diffeomorphism invariance across this extended structure.

The para-Hermitian structure introduced above--defined by a product structure $K$ satisfying $K^2 = \mathbb{I}$ and a neutral metric $\eta$ obeying $\eta(KX, KY) = -\eta(X, Y)$-is closely related to the framework of generalized complex geometry developed by Gualtieri~\cite{Gualtieri2004}. In generalized complex geometry, the phase space is extended from the tangent bundle $TM$ to the generalized tangent bundle $TM \oplus T^*M$, equipped with a natural pairing of signature $(n,n)$:
\begin{equation}
\langle X + \xi, Y + \eta \rangle = \frac{1}{2} \left( \xi(Y) + \eta(X) \right),
\end{equation}
where $X, Y \in TM$ and $\xi, \eta \in T^*M$. This inner product plays the role of the neutral metric $\eta$ in the para-Hermitian context.

A generalized complex structure $\mathcal{J}$ is then a bundle map $\mathcal{J}: TM \oplus T^*M \to TM \oplus T^*M$ satisfying:
\begin{equation}
\mathcal{J}^2 = -\mathbb{I}, \qquad \langle \mathcal{J}u, \mathcal{J}v \rangle = \langle u, v \rangle,
\end{equation}
and integrability defined with respect to the Courant bracket. In contrast, a para-complex structure replaces $\mathcal{J}$ with an endomorphism $K$ satisfying $K^2 = \mathbb{I}$ and $\eta(Ku, Kv) = -\eta(u, v)$, distinguishing it as a real structure. This makes para-Hermitian geometry a real analog of generalized complex geometry, suitable for describing doubled or bi-temporal structures such as the $T^2$ manifold of complex time. The decomposition of the bundle into eigenbundles of $K$,
\begin{equation}
TM \oplus T^*M = L_+ \oplus L_-,
\end{equation}
where $L_\pm$ are the $\pm 1$ eigenbundles of $K$, mimics the decomposition of generalized complex manifolds into holomorphic and antiholomorphic structures, but now adapted to time-like or neutral signature applications. This correspondence allows the formalism of temporal $T$-duality, doubled geometry, and non-commutative time to be expressed in a covariant language that unifies causality and coherence across the temporal manifold. It also lays the groundwork for embedding the $T^2$ framework into generalized flux backgrounds and non-geometric string compactifications.

The generalized metric takes the form:
\begin{equation}
    \mathcal{G} = \begin{pmatrix} g - B g^{-1} B & B g^{-1} \\ -g^{-1} B & g^{-1} \end{pmatrix},
\end{equation}
where $B$ is an antisymmetric two-form encoding non-commutativity and coherence effects. From this structure, the decoherence functional $\Gamma[\phi]$ can be understood as arising from $B$-field couplings:
\begin{equation}\label{eq:Gamma_phi}
    \Gamma[\phi] = \int_{T^2} dt_1 dt_2 \, ( \gamma_0 \dot{\phi}^2 + \gamma_1 \phi^2 ) + \int d^4x \, G(x - y) J(y),
\end{equation}
with environmental influence encoded in $J(y)$ and propagation across $t_2$ in the Green's function $G$. This formalism mirrors the structure of generalized complex geometry introduced by Gualtieri~\cite{Gualtieri2004}, where the endomorphism $J$ satisfies $J^2 = -\mathbb{I}$ and is compatible with a natural pairing on $TM \oplus T^*M$. The para-Hermitian structure used here instead features a real product structure $K$ with $K^2 = +\mathbb{I}$, enabling a bi-temporal foliation aligned with the causal (t$_1$) and coherence (t$_2$) directions of the manifold $T^2$. This allows the metric, B-field, and decoherence functional $\Gamma[\phi]$ to be expressed covariantly in the language of generalized flux backgrounds.

This formulation unifies causal and coherent dynamics, and sets the stage for the introduction of gauge symmetry and quantum gravity in the subsequent sections.

\section{Quantum Dynamics and Measurement}
In conventional quantum mechanics, measurement is typically modeled as an abrupt, non-unitary collapse of the wavefunction, standing in contrast to the continuous, unitary evolution governed by the Schr\"odinger equation. In the $T^2$ framework, however, this dichotomy is resolved geometrically. Coherence and measurement are encoded in the two orthogonal temporal directions, $t_1$ and $t_2$, and collapse becomes a projection across the temporal manifold rather than a fundamental discontinuity~\cite{zurek2003decoherence, everett1957relative}.

Quantum evolution in $T^2$ proceeds according to the extended Schr\"odinger equation~\eqref{eq:GSE} with unitary evolution along $t_1$ and coherence decay along $t_2$. This naturally yields solutions with exponential damping of coherence:
\begin{equation}
    \Psi(\vec{r}, t_1, t_2) = \psi(\vec{r}) e^{i k_1 t_1} e^{-k_2 t_2}, \quad |\Psi|^2 \sim e^{-2k_2 t_2}.
\end{equation}

To illustrate how coherence time $t_2$ modulates interference in the two-time framework, consider a quantum superposition of two energy eigenstates with amplitudes $\alpha$ and $\beta$. The general form of such a state on $T^2$ evolves as:
\begin{equation}
\Psi(t_1, t_2) = \alpha \Psi_1 e^{-i E_1 t_1 / \hbar} e^{-E_1 t_2 / \hbar} + \beta \Psi_2 e^{-i E_2 t_1 / \hbar} e^{-E_2 t_2 / \hbar},
\end{equation}
then interference terms in $|\Psi|^2$ acquire the phase
\begin{equation}
e^{i (E_2 - E_1) t_2 / \hbar} = e^{i \Delta E t_2 / \hbar},
\end{equation}
reflecting the spectral structure of coherence. This factor modulates the interference pattern and links $t_2$ to the energy separation $\Delta E$, supporting the interpretation of $t_2$ as a direction of phase coherence. This demonstrates how interference visibility degrades in time as coherence diminishes, offering a geometric account of decoherence consistent with environmental entanglement.

This structure generalizes the von Neumann projection postulate. Rather than treating collapse as an ad hoc operation, it is recast as a boundary condition on a two-dimensional temporal sheet. The full mixed-state dynamics is governed by a two-time Lindblad equation:
\begin{equation}
    \frac{\partial \rho}{\partial t_1} + i \frac{\partial \rho}{\partial t_2} = -i[\hat{H}, \rho] + \mathcal{L},
\end{equation}
where $\mathcal{L}$ is a Lindblad superoperator encoding environmental interactions and decoherence~\cite{breuer2002theory, gisin1992quantum}. 

This equation arises naturally from a partial trace over environmental degrees of freedom coupled to $t_2$, linking the formal structure to open-system quantum dynamics~\cite{joos2003decoherence, caldeira1983path}. Measurement, in this picture, corresponds to the embedding of a Cauchy surface in $T^2$, where quantum amplitudes are projected onto a definite branch across the coherence direction.

Interference and collapse are modulated by phase factors in $t_2$, with the relative phase between two states $\Psi_1$ and $\Psi_2$ evolving as $\exp(i \Delta E t_2 / \hbar)$. The time-averaged probability for a superposition,
\begin{equation}\label{eq:pdf_sup}
    |\Psi|^2 = |\alpha|^2 |\Psi_1|^2 + |\beta|^2 |\Psi_2|^2 + 2\text{Re}[\alpha^* \beta \Psi_1^* \Psi_2 e^{i \Delta E t_2 / \hbar}],
\end{equation}
reveals that $t_2$ governs observable interference patterns. Decoherence corresponds to a damping of this cross-term, resulting in effective classicality~\cite{schlosshauer2005decoherence, zurek2003decoherence}. The exponential factor $e^{i \Delta E t_2 / \hbar}$ in the expression,~\eqref{eq:pdf_sup} arises from the relative phase evolution between two energy eigenstates along the coherence axis $t_2$. In the extended two-time formalism, each component $\Psi_i$ evolves as $e^{- E_i t_2 / \hbar}$ in the compactified coherence direction. When computing the interference term in the modulus squared of the total wavefunction, the difference in these exponential phases results in the factor $e^{i \Delta E t_2 / \hbar}$, where $\Delta E = E_2 - E_1$. This term captures the modulation of quantum interference as a function of coherence time and encodes spectral information in the evolution of superposed states.
The evolution of quantum coherence over the temporal manifold can be described by a diffusion equation of the form:
\begin{equation}
\left( \frac{\partial}{\partial t_2} + \gamma \right) D(t_1, t_2) = D \frac{\partial^2}{\partial t_1^2} D(t_1, t_2),
\end{equation}
where $D(t_1, t_2)$ denotes the decoherence function, $D$ is a diffusion coefficient, and $\gamma$ encodes the decay rate due to system-environment coupling.

This structure captures non-Markovian behavior and coherence revivals. The environmental response enters via the spectral density and decoherence kernel, resulting in temporal coherence functions of the form:
\begin{equation}
f(t_1) = \exp\left[ -\left( \frac{t_1}{\tau_{\mathrm{env}}} \right)^\alpha \right], \quad 1 < \alpha \leq 2,
\end{equation}
where $\alpha$ characterizes memory effects and $\tau_{\mathrm{env}}$ is the environmental timescale.
The role of $t_2$ as a coherence axis parallels modular flow in quantum field theory and holography, where thermal time arises from the structure of entanglement and modular Hamiltonians~\cite{faulkner2014modular, cardy2016scaling}. In this analogy, the measurement surface is a modular slice of the temporal manifold, selecting a particular pointer basis. Such slicing is covariant within the para-Hermitian geometry of $\mathbb{T}M$, ensuring that quantum state reduction respects generalized diffeomorphism symmetry. Altogether, the two-time formalism reinterprets measurement and collapse not as axiomatic postulates, but as geometric operations emerging from the internal structure of time itself. This offers an unification of unitary dynamics, decoherence, and observation under the $T^2$ manifold, while remaining consistent with both open-system quantum theory and gravitational dualities.

\section{Geometry and Classical Emergence}
Entanglement traditionally refers to correlations between spatially separated quantum systems, but in the $T^2$ framework, such correlations extend across the two temporal directions $t_1$ and $t_2$. This structure reveals a richer class of entangled states, where quantum coherence and causal propagation become geometrically encoded within the manifold of time itself.

Consider two subsystems $A$ and $B$ evolving jointly on $T^2$, with an entangled state represented as:
\begin{equation}
\Psi(t_1, t_2) = \frac{1}{\sqrt{2}} \left[ \psi_A(t_1, t_2) \otimes \varphi_B(t_1, t_2) + \varphi_A(t_1, t_2) \otimes \psi_B(t_1, t_2) \right],
\end{equation}
where interference is preserved along both temporal axes. The corresponding two-time correlation function,
\begin{equation}
C(t_1, t_2; t_1', t_2') = \langle \Psi | \hat{A}(t_1, t_2) \otimes \hat{B}(t_1', t_2') | \Psi \rangle,
\end{equation}
captures not only causal relations but also coherence propagation across temporal slices. This expands the conventional framework of quantum nonlocality into higher-dimensional time.
These structures enable a temporal generalization of the Bell–CHSH inequality. Define a temporal correlator $S$ as:
\begin{align}
S = \frac{1}{2} \Big[ C(t_1, t_2; t_1', t_2') + C(t_1, t_2; t_1', t_2'')  + C(t_1'', t_2''; t_1', t_2') - C(t_1'', t_2''; t_1', t_2'') \Big],
\end{align}
which satisfies $|S| \leq 2$ under local realism, but achieves maximal quantum violation $S = 2\sqrt{2}$ when phase alignment along $t_2$ satisfies:
\begin{equation}
\frac{\Delta E}{\hbar} \Delta t_2 = \frac{\pi}{4} + n\pi, \quad n \in \mathbb{Z},
\end{equation}
as predicted in modular Hamiltonian frameworks \cite{faulkner2014modular, cardy2016scaling}.

To formalize decoherence in this setting, the density matrix must incorporate coherence decay in both directions:
\begin{equation}
\rho(t_1, t_2; t_1', t_2') = \ket{\Psi(t_1, t_2)}\bra{\Psi(t_1', t_2')} \exp\left[ -\frac{(t_1 - t_1')^2}{2\tau_1^2} - \frac{(t_2 - t_2')^2}{2\tau_2^2} \right],
\end{equation}
where $\tau_1$ and $\tau_2$ are coherence timescales. The exponential damping describes temporal decoherence consistent with Lindblad evolution \cite{zurek2003decoherence, caldeira1983path}.

The direction $t_2$ functions as a modular parameter akin to the Euclidean thermal circle in field theory, connecting this formalism to holographic dualities. Winding around the compact $t_2$ cycle encodes entanglement structure in ways analogous to replica wormhole configurations in AdS/CFT \cite{almheiri2020replica}. As such, $t_2$ governs not only coherence but also the entropic architecture of spacetime.

The classical limit emerges naturally in this framework via projection onto $t_1$ slices. Quantum states, initially delocalized along $t_2$, decohere into pointer states as environmental interactions suppress interference. The reduced classical density matrix arises by integrating out the $t_2$ direction:
\begin{equation}
\rho_{\text{classical}}(t_1) = \int dt_2 \, \rho(t_1, t_2),
\end{equation}
localizing probability in $t_1$ and recovering classical determinism.

The emergent classicality can also be viewed through the lens of phase space coarse-graining. Temporal uncertainty and non-commutativity between $t_1$ and $t_2$ imply a minimal resolution volume, analogous to spatial phase space cells:
\begin{equation}
    \Delta t_1 \Delta t_2 \geq \frac{\hbar}{2E},
\end{equation}
which suppresses fine-scale interference patterns and enables classical trajectories to form. This projection process is equivalent to a temporal coarse-graining or time-averaging, which is not imposed externally but emerges from the internal geometry of the temporal manifold. The classical limit then corresponds to trajectories on the quotient space $T^2 / U(1)$, where the $U(1)$ symmetry reflects the periodicity or gauge redundancy in $t_2$. Furthermore, in systems with semiclassical limits, the action integral over complex time contours becomes stationary not merely with respect to $t_1$ but with respect to variations in both $t_1$ and $t_2$. The classical equations of motion thus arise as extrema of a generalized action principle on $T^2$, thereby surviving the averaging process.

The dynamics of quantum fields on $T^2$ can be derived from a generalized variational principle:
\begin{equation}
\delta \int_{T^2} \left( L - i \Gamma \right) dt_1 dt_2 = 0, \label{eq:variational_t2}
\end{equation}
where $L$ is the classical Lagrangian and $\Gamma$ encodes coherence dissipation due to compactification or environmental interactions. This principle selects paths that extremize the classical action while minimizing decoherence, yielding trajectories that are both dynamically viable and stable under projection to $t_1$. In this sense, measurement, collapse, and classical determinism emerge from geometric and analytic features of complexified time. The two-time formalism thus not only resolves interpretational ambiguities but elevates the geometry of time itself to a central dynamical agent in the quantum-classical interface.



\section{Quantum Gravity and Minimal Temporal Structure}
The compactification of the coherence coordinate $t_2$ introduces a fundamentally new ingredient into the structure of quantum gravity. Much like Kaluza-Klein compactifications in higher-dimensional unification, compactifying time rather than space leads to quantized temporal modes, dualities, and UV regulators that govern coherence loss and gravitational entropy. This section explores how minimal temporal structure, encoded in the geometry of $T^2$, alters our understanding of spacetime, black holes, and string theory.

In the context of string theory, the worldsheet action admits natural extension to include a compactified temporal direction:
\begin{equation}
S = -\frac{1}{4\pi \alpha'} \int d^2\tau \, \sqrt{-h} \, h^{ab} \, \partial_a X^\mu \partial_b X_\mu,
\end{equation}
with a compactified time direction obeying
\begin{equation}
X^{t_2}(\tau_1, \tau_2 + 2\pi R_{t_2}) = X^{t_2}(\tau_1, \tau_2) + 2\pi w R_{t_2}.
\end{equation}
This structure leads to two distinct quantized spectra:
\begin{align}\label{eq:quntized_spectra}
    p_{t_2} = \frac{n \hbar}{R_{t_2}}, \qquad E_w = \frac{w^2 R_{t_2}^2}{\alpha'}\end{align}
These are exchanged under T-duality,
\begin{equation}
    R_{t_2} \longleftrightarrow \frac{\alpha'}{R_{t_2}},
\end{equation}
leading to a minimal temporal resolution $\Delta t_2 \sim \sqrt{\alpha'}$, analogous to spatial quantization in loop quantum gravity~\cite{strominger1996microscopic, rovelli1998loop}. While T-duality is most familiar in compactified spatial dimensions, its application to time—particularly in Euclideanized formulations and thermal dualities—is well supported in string theory~\cite{giveon1994target, atick1988hagedorn}.

This minimal coherence scale regularizes UV divergences in quantum field theory. For instance, the two-point correlation function acquires a Gaussian regulator in $t_2$:
\begin{equation}
    G(t_1, t_2) \sim e^{-t_2^2/4\alpha'},
\end{equation}
which suppresses short-distance divergences along the coherence axis. In the presence of a background $B$-field, the effective worldsheet theory becomes noncommutative in time:
\begin{equation}
    [t_1, t_2] = i \theta \alpha',
\end{equation}
resulting in a modified uncertainty principle:
\begin{equation}
    \Delta t_1 \, \Delta t_2 \geq \frac{\theta \alpha'}{2},
\end{equation}
and operator-valued temporal metrics:
\begin{equation}
    ds^2 = \text{Tr}[g_{\mu\nu}(t_1, t_2)] \, dt^\mu dt^\nu + \mathcal{O}(\theta^2 \alpha'^2).
\end{equation}
This bracket structure supports UV/IR mixing and regulates short-distance divergences~\cite{seiberg1999string, sazabo2003quantum}.

In gravitational settings, compactified $t_2$ modifies the Einstein equations. The curvature response is governed by coherence-averaged energy density:
\begin{equation}
    R_{\mu\nu} - \frac{1}{2} R g_{\mu\nu} = \frac{8\pi G}{c^4} \langle \hat{T}_{\mu\nu} \rangle_{t_2},
\end{equation}
with $\langle \cdot \rangle_{t_2}$ denoting integration over the coherence cycle. This framework allows consistent inclusion of weak measurements, quantum backreaction, and spacetime correlations.

To ensure gauge symmetry is preserved in the presence of coherence evolution, we extend BRST quantization to include $t_2$-dependent ghost and anti-ghost fields. Let $c^a(t_1, t_2)$, $\bar{c}_a(t_1, t_2)$, and $b^a(t_1, t_2)$ be ghost, anti-ghost, and Nakanishi--Lautrup fields, respectively. The BRST transformations generalize to:
\begin{align}
    \mathcal{Q} A_\mu &= D_\mu c, \\
    \mathcal{Q} c &= -\frac{1}{2}[c, c], \\
    \mathcal{Q} \bar{c} &= b, \\
    \mathcal{Q} b &= 0.
\end{align}
The gauge-fixed BRST-invariant action becomes
\begin{equation}
    S_{\text{BRST}} = S_{\text{classical}} + \mathcal{Q} \int_{T^2} \text{Tr}(\bar{c} \, \mathcal{F}[A]) dt_1 dt_2,
\end{equation}
with $\mathcal{F}[A]$ the gauge-fixing function. The full path integral is then
\begin{equation}
    Z = \int \mathcal{D}A \, \mathcal{D}c \, \mathcal{D}\bar{c} \, \mathcal{D}b \, e^{i S_{\text{BRST}}},
\end{equation}
ensuring that gauge invariance and unitarity are preserved under complexified evolution.

The dynamics over the two-time manifold $T^2$ can be captured by a generalized action principle that incorporates both unitary propagation and coherence decay. Instead of a standard time-evolution functional, we consider a path integral weighted by a complexified action:
\begin{equation}
    S_{T^2}[\phi] = \int_{T^2} dt_1 dt_2 \left( \mathcal{L}[\phi] - i \Gamma[\phi] \right),
\end{equation}
where $\mathcal{L}[\phi]$ is the Lagrangian governing unitary evolution along $t_1$, and $\Gamma[\phi]$ is a real-valued functional describing decoherence and dissipation along the coherence axis $t_2$~\cite{feynman1963theory, caldeira1983path}.

The variational principle applied to $S_{T^2}$ yields equations of motion that are extremal not only in the causal direction but also with respect to the coherence flow:
\begin{equation}
    \delta S_{T^2} = 0 \quad \Rightarrow \quad \frac{\delta \mathcal{L}}{\delta \phi} = i \frac{\delta \Gamma}{\delta \phi}.
\end{equation}
This equation defines a balance condition between classical propagation and coherence decay, ensuring that physical trajectories survive under projection to classical $t_1$-slices. In semiclassical limits, this reproduces the decoherence-modified classical dynamics observed in pointer basis emergence and environment-induced superselection~\cite{zurek2003decoherence, joos2003decoherence}.

Furthermore, when applied to semiclassical black hole systems, this generalized action principle provides a framework for computing saddle-point contributions to path integrals that include both thermal and coherence information. This supports a unified description of black hole evaporation, information recovery, and entropy production in the presence of compactified time.

Finally, black hole thermodynamics and temporal discretization reveal rich structure. The Euclideanized near-horizon metric becomes cigar-shaped:
\begin{equation}
    ds^2 = \frac{d\tau_2^2}{1 - e^{-(\tau_2/\Delta t_2)^2}} + \left(1 - e^{-(\tau_2/\Delta t_2)^2}\right) d\varphi^2,
\end{equation}
resolving the conical singularity at $\tau_2 = 0$. The black hole area spectrum is quantized:
\begin{equation}
    A_n = 4\pi \alpha' (2n + 1),
\end{equation}
and the entropy becomes
\begin{equation}
    S_{\text{BH}} = \frac{k_B A}{4 \ell_P^2} - \frac{3}{2} k_B \log \left( \frac{A}{\ell_P^2} \right) + \mathcal{O}(\ell_P^2/A).
\end{equation}
Altogether, the $T^2$ framework weaves together quantum geometry, duality, coherence, and gravity into a consistent two-time picture of minimal temporal structure.

\section{Holography and Quantum Information}
The compactification of the coherence coordinate $t_2$ not only plays a foundational role in quantum gravity but also bridges several threads of modern physics—holography, entropy bounds, quantum information, and modular dynamics. The structure of $T^2$ supports a dual interpretation of time in terms of causal propagation and coherence evolution, yielding a natural framework for encoding entanglement and recovering unitarity through geometric constraints.

In the AdS/CFT correspondence, compactification of the coherence direction $t_2$ introduces winding sectors and modular transformations that restructure the entanglement properties of the boundary theory. For instance, the entropy of a 2D conformal field theory acquires a duality-invariant form:
\begin{equation}
S = \frac{\pi^2 c}{3} \left( \frac{R_{t_2}}{\beta} + \frac{\beta}{R_{t_2}} \right),
\end{equation}
which is symmetric under the exchange $R_{t_2} \leftrightarrow \beta$. Identifying the compactification radius with the Schwarzschild radius, $R_{t_2} \sim 2GM/c^2$, reproduces the Bekenstein-Hawking entropy:
\begin{equation}
S = \frac{A}{4 G \hbar},
\end{equation}
revealing that gravitational entropy can emerge from the geometry of compactified coherence time~\cite{cardy1986operator, strominger1996microscopic}.

This duality structure is deeply linked to modular flow in quantum field theory. In conformal settings, the modular Hamiltonian $K$ generates unitary evolution along entanglement-preserving surfaces and takes the form:
\begin{equation}
K = 2\pi \int_\Sigma x^1 T_{00}(x) \, d^{d-1}x, \label{eq:modular_generator}
\end{equation}
where $\Sigma$ is the entangling surface and $T_{00}(x)$ is the energy density operator. This structure naturally arises in the Rindler wedge and underlies the thermodynamics of entanglement entropy in holographic systems~\cite{faulkner2014modular, jafferis2016relative}.

The two-time formalism generalizes this construction. The compact coherence direction $t_2$ plays a modular role, allowing us to interpret evolution along complex time in terms of modular flow. A generalized modular evolution operator becomes:
\begin{equation}
U(\tau) = e^{-i K \tau / \hbar}, \label{eq:modular_evolution}
\end{equation}
encapsulating both causal dynamics (via $t_1$) and coherence decay (via $t_2$). Its spectral decomposition encodes the entanglement structure across slices of modular time, supporting a geometric interpretation of information flow. Furthermore, the reduced state $\rho_A$ of a subregion admits a modular decomposition:
\begin{equation}
\rho_A = \frac{1}{Z} \exp(-K_A),
\end{equation}
where $K_A$ is the modular Hamiltonian generating modular flow. In the $T^2$ framework, this generator extends to:
\begin{equation}
K(\tau) = \int_{\Sigma(\tau)} d\Sigma^\mu \, T_{\mu\nu} \, \zeta^\nu(\tau),
\end{equation}
with $\zeta^\nu(\tau)$ a modular flow vector defined over the complexified temporal manifold~\cite{witten2018aps, lewkowycz2013generalized}. These modular structures find further support in Wilson loop amplitudes, which acquire temporal winding corrections in the presence of compact $t_2$:
\begin{equation}
\langle w_{t_2} | W_\gamma \rangle \sim \exp\left( - S_{\text{EH}}[\gamma] \right), \quad \langle W_\gamma \rangle \sim \exp\left( - \frac{\text{Area}(\gamma)}{\alpha'} \right),
\end{equation}
suggesting that winding along $t_2$ generates microscopic Einstein–Rosen bridges in line with the ER=EPR correspondence~\cite{maldacena2013cool, almheiri2020replica}. This construction generalizes the role of modular Hamiltonians beyond spatial subregions to include slices of complexified time. For instance, in a thermal state with modular parameter $t_2 \sim i\beta$, the operator $K(\tau)$ generates evolution along entanglement-preserving flows. In the Rindler wedge, this corresponds to boosts in Minkowski space, while in AdS/CFT, modular flow is dual to radial evolution in the bulk. The compactification of $t_2$ introduces discrete modular spectra, tightly linked to replica wormhole contributions and entropy regulation.

On the worldsheet, the string path integral over $T^2$ includes both momentum and winding excitations, see eq.~\eqref{eq:quntized_spectra}:
\begin{equation}
    X^{t_2}(\tau + 2\pi) = X^{t_2}(\tau) + 2\pi w R_{t_2}.
\end{equation}

Thermal partition functions on compact time geometries are modular-invariant and UV-finite. The geometry of the Euclidean manifold becomes cigar-shaped:
\begin{equation}
    ds^2 = \frac{d\tau_2^2}{1 - e^{-(\tau_2/\Delta t_2)^2}} + \left(1 - e^{-(\tau_2/\Delta t_2)^2} \right) d\phi^2,
\end{equation}
regularizing conical singularities and suppressing high-energy modes. The path integral measure becomes
\begin{equation}
    \mathcal{D}X \rightarrow \mathcal{D}X \prod_n \left(1 - e^{-(2\pi n \Delta t_2 / \beta)^2} \right)^{D-2},
\end{equation}
introducing modular suppression of UV divergences.

The Hagedorn density of states is tamed by this structure:
\begin{equation}
    \rho(E) \sim e^{\beta_H E} \exp\left(-\frac{E^2 \alpha'}{4 \ln(E/E_0)}\right),
\end{equation}
softening the divergence and stabilizing entropy at high energy.

Black hole horizons exhibit discrete radial displacements:
\begin{equation}
    \delta r \sim \sqrt{\alpha'} \left(n + \tfrac{1}{2} \right),
\end{equation}
leading to corrected Hawking temperature:
\begin{equation}
    T_H = T_0 \left[ 1 + \frac{\pi^2 \alpha'}{4 r_s^2}\left(n^2 + n + \tfrac{1}{2} \right) \right]^{-1},
\end{equation}
and logarithmic entropy corrections:
\begin{equation}
    S_{\text{BH}} = \frac{k_B A}{4 \ell_P^2} - \frac{3}{2} k_B \ln \left( \frac{A}{\ell_P^2} \right) + \mathcal{O}(\ell_P^2 / A).
\end{equation}
The compactified coherence scale thus regulates high-frequency behavior, entropy growth, and horizon microstates.

Replica wormhole contributions to the gravitational path integral are suppressed geometrically:
\begin{equation}
    Z_n \sim \exp\left( - \frac{n \Delta t_2}{G_N} S_{\text{bulk}} \right),
\end{equation}
leading to a radiation entropy curve:
\begin{equation}
    S_{\text{rad}}(t) = S_{\text{th}} \left( 1 - e^{-t^2/\Delta t_2^2} \right) + \frac{\pi c}{6} \frac{\Delta t_2}{\beta} \, \text{erf}\left( \frac{t}{\Delta t_2} \right),
\end{equation}
and restoring unitarity at late times via compact temporal regularization. The modular Hamiltonian $\hat{K}$ associated with a subregion in the boundary CFT can be interpreted geometrically as a boost generator conjugate to the compactified coherence direction $t_2$. In analogy with entanglement wedge dynamics, it takes the form:
\begin{equation}
\hat{K} = \int_{\Sigma} \xi^\mu T_{\mu\nu} d\Sigma^\nu,
\end{equation}
where $\xi^\mu$ is a modular flow vector field and $T_{\mu\nu}$ is the boundary stress-energy tensor~\cite{faulkner2014modular, jafferis2016relative}. For thermalized subsystems with coherence periodicity $t_2 \sim i \beta$, the modular Hamiltonian governs both entanglement entropy and the modular time evolution. This structure allows winding along $t_2$ to encode emergent causal structure in holographic duals.

In JT gravity, the wormhole-modified partition function becomes:
\begin{equation}
    Z_{\text{wormhole}} = \int_0^\infty \frac{dE}{2\pi} \rho(E) e^{-n E \Delta t_2} |\langle E | \psi \rangle|^2,
\end{equation}
serving as an infrared regulator and entropic cap on late-time dynamics.

Taken together, these results show that the compactified coherence direction $t_2$ encodes modular symmetry, suppresses chaos, discretizes entropy, and unifies thermodynamics with information flow. The $T^2$ framework thus renders the structure of unitarity and holography geometric and exact.

\section{Experimental Consequences}
Although the compactified coherence dimension $t_2$ is not directly observable, its physical consequences pervade multiple domains of high-precision measurement and astrophysical observation. The key experimental signature is the existence of a minimal coherence scale $\Delta t_2$, which induces quantized temporal modes, dispersion relations, and non-commutative structure. These lead to observable consequences in ultrafast optics, high-energy particle astrophysics, and black hole thermodynamics.

A particularly promising observational signature lies in the spectral modulation induced by the compactified coherence direction $t_2$. Temporal winding and quantized modes introduce fine structure into radiation and propagation spectra, especially for systems sensitive to phase coherence. This effect manifests in two primary ways: first, as frequency sidebands in ultrafast optical measurements with spacing
\begin{equation}
    \Delta \omega \sim \frac{c}{\sqrt{\alpha'}} \sim 10^{21} \, \text{Hz}, 
\end{equation}
and second, as energy-dependent arrival-time dispersion for high-energy astrophysical messengers. For instance, gamma-ray bursts exhibit delays of the form
\begin{equation}
    \frac{\Delta t}{t} \sim \frac{E^2}{M_s^2 c^4},
\end{equation}
while modified dispersion relations for neutrinos take the form
\begin{equation}
    E^2 = p^2 c^2 + m^2 c^4 + \xi \frac{E^4}{M_s^2 c^4}.
\end{equation}
These imprints of the $t_2$-geometry offer empirical access to quantum coherence structure and motivate dedicated searches in high-precision timing experiments, astrophysical observatories, and particle detectors. This structure aligns with the thermal time hypothesis~\cite{connes1994noncommutative, rovelli1993statistical}, where modular flow defines intrinsic evolution in generally covariant systems. The coherence direction $t_2$ acts as a modular parameter, and thermalization emerges geometrically from entanglement across compactified temporal cycles.

In ultrafast quantum optics, the irreducible temporal resolution is set by
\begin{equation}
    \delta t \sim \frac{\sqrt{\alpha'}}{c} \sim 10^{-19} \, \text{s},
\end{equation}
producing temporal jitter in attosecond-scale pump-probe measurements. The resulting modulation of frequency spectra generates sidebands of order
\begin{equation}
    \Delta \omega \sim \frac{c}{\sqrt{\alpha'}} \sim 10^{21} \, \text{Hz},
\end{equation}
manifesting as phase decoherence.

In astrophysical propagation, coherence-scale effects lead to energy-dependent time delays. For gamma-ray bursts (GRBs), the accumulated dispersion over cosmological distances yields
\begin{equation}
    \frac{\Delta t}{t} \sim \frac{E^2}{M_s^2 c^4},
\end{equation}
placing strong bounds on $M_s$ from Fermi-LAT data, with constraints $M_s \gtrsim 5 \times 10^{17}$ GeV. Neutrino-photon coincidence measurements from GRB 170817A impose similar bounds via the deformed dispersion relation:
\begin{equation}
    E^2 = p^2 c^2 + m^2 c^4 + \xi \frac{E^4}{M_s^2 c^4},
\end{equation}
where IceCube and Super-Kamiokande limit $\xi < 10^{-3}$.

Compactification also alters thresholds for high-energy processes. In the Greisen-Zatsepin-Kuzmin (GZK) cutoff for ultra-high-energy cosmic rays (UHECRs), quantized temporal momenta modify energy balance:
\begin{equation}
    E^2 = p^2 c^2 + m^2 c^4 + \left( \frac{n \hbar c}{R_{t_2}} \right)^2,
\end{equation}
shifting the pion production threshold:
\begin{equation}
    E_{\text{th}} = \frac{m_\pi c^2 (2 m_p c^2 + m_\pi c^2)}{4 \epsilon_\gamma} \left[ 1 + \frac{n^2 \hbar^2 c^2}{R_{t_2}^2 m_p^2 c^4} \right].
\end{equation}
Data from the Pierre Auger Observatory set bounds $R_{t_2} < 0.7 \times 10^{-19}$ meters.

Compact time also impacts baryon decay, providing exponential suppression:
\begin{equation}
    \Gamma_{p \rightarrow n} \sim \exp\left( - \frac{R_{t_2} \Delta m c}{\hbar} \right),
\end{equation}
helping to account for the exceptional stability of protons. The absence of observed proton decay imposes lower bounds on $M_s$ and coherence length.

Collectively, these results constrain $\Delta t_2$ to lie near the string scale. Quantum optics, neutrino telescopes, gravitational wave interferometry, and cosmic ray observatories all provide platforms where coherence-induced deviations may emerge. The universality of these effects supports the physical reality of the compact $t_2$ dimension. Although indirect, their signatures are concrete: phase decoherence, arrival-time shifts, spectrum modifications, and threshold anomalies. As observational capabilities improve, they offer a compelling window into the geometry of time.

\section{Theoretical Unification and Emergent Temporal Geometry}
The compact temporal manifold $T^2 = (t_1, t_2)$ offers a unifying structure that reconciles coherence dynamics, causal evolution, and thermodynamic entropy within a single geometric framework. This dual-time description embeds both modular flow and gravitational degrees of freedom, allowing phenomena across quantum theory, general relativity, and holography to be understood as emergent from a compactified temporal substrate.

In the thermal time hypothesis, the arrow of time is linked to the modular flow of quantum states. Within the $T^2$ geometry, inverse temperature is encoded as the length of the compactified coherence cycle, leading to a geometric interpretation:
\begin{equation}
    \beta = \frac{2\pi \Delta t_2}{\ln(1 + \sqrt{2})},
\end{equation}
where $\Delta t_2$ represents the minimal coherence interval. Thermality is no longer an imposed boundary condition but an emergent feature from the temporal geometry.

Information recovery in black hole evaporation also follows naturally from this compact temporal structure. Replica wormholes, central to the gravitational path integral in recent semiclassical models, are suppressed by coherence scale effects:
\begin{equation}
    \chi_n \sim e^{-n \Delta t_2 / G_N},
\end{equation}
ensuring a Page-curve-like saturation of entanglement entropy:
\begin{equation}
    S_{\text{ent}}(t) = S_{\text{BH}} \left(1 - e^{-t / \Delta t_2} \right).
\end{equation}
This smooth evolution avoids firewalls or discontinuities, embedding unitary information dynamics into the compact $t_2$ geometry.

Quantum chaos is likewise moderated by coherence quantization. The Lyapunov exponent governing exponential divergence in chaotic systems is corrected:
\begin{equation}
    \lambda_L = \frac{2\pi}{\beta} \left(1 - \left( \frac{\Delta t_2}{\beta} \right)^2 \right),
\end{equation}
indicating suppressed scrambling at high energies. Simultaneously, the number of accessible quantum states is bounded by coherence resolution:
\begin{equation}
    N = \frac{e^{S_{\text{BH}}}}{\Delta t_2^2},
\end{equation}
suggesting a compact-time limit to entropy and complexity.

In loop quantum gravity (LQG), a two-time Wheeler--DeWitt equation naturally arises:
\begin{equation}
    \left( \frac{\delta^2}{\delta t_1^2} + \frac{\delta^2}{\delta t_2^2} - \hat{H}_{\text{geo}} \right) |\Psi(t_1, t_2)\rangle = 0,
\end{equation}
where $\hat{H}_{\text{geo}}$ includes quantum gravitational curvature. The area gap of LQG imposes a temporal uncertainty:
\begin{equation}
    \Delta t_1 \, \Delta t_2 \geq \ell_P^2,
\end{equation}
supporting discrete, quantized temporal evolution.

The effective Hamiltonian itself receives coherence-dependent corrections:
\begin{equation}
    \hat{H}_{\text{geo}} = \hbar^2 \left[ \gamma^2 \hat{R} + \frac{\Lambda}{3} + \alpha \left( \frac{\delta}{\delta t_2} \right)^2 \hat{R} \right],
\end{equation}
where $\gamma$ is the Barbero--Immirzi parameter and $\alpha$ a coupling to coherence curvature. These corrections may regularize singularities and allow backreaction from $t_2$ fluctuations.

Coherence dynamics along $t_2$ may be gapless or gapped depending on the phase:
\begin{align}
    \omega_{t_2}^2 &= v^2 k_{t_2}^2, \quad \langle O(t_2) O(0) \rangle \sim |t_2|^{-\Delta} \, &\text{(gapless)} \\
    \omega_{t_2}^2 &= v^2 k_{t_2}^2 + m_{t_2}^2, \quad \langle O(t_2) O(0) \rangle \sim e^{-m_{t_2} |t_2|} \, &\text{(gapped)}
\end{align}
representing a coherence phase structure similar to renormalization group flow in field theory.

Finally, the AdS/CFT correspondence links $t_2$ to scaling dimensions of boundary operators:
\begin{equation}
    \Delta t_2 = \ell_{\text{AdS}} \exp\left( -\frac{\pi c}{3\Delta} \right),
\end{equation}
establishing a duality between coherence localization and boundary entanglement.

The compactified temporal dimension thus controls thermal behavior, gravitational entropy, quantum chaos, and renormalization flow—revealing that geometry, coherence, and information are unified in the structure of time itself.

\section{Ontological and Philosophical Implications}
The $T^2$ framework not only restructures the technical foundations of quantum gravity and holography, but also reconfigures long-standing ontological views of time and measurement. Traditionally, the debate over presentism versus eternalism has framed time as either dynamically emergent or statically extended. However, the introduction of a second temporal dimension---coherence time $t_2$---permits a novel pluralist interpretation: classical causality evolves along $t_1$, while quantum coherence and branching unfold along $t_2$.

In this view, measurement is not a stochastic projection at a moment in time but a geometric slice across $T^2$ that selects a decohered branch. Temporal locality is defined not solely in $t_1$, but via projections in both temporal directions. This dual structure dissolves the tension between unitary evolution and wavefunction collapse, offering a continuous geometric basis for quantum measurement.

The probability amplitude must now be defined over the full temporal manifold. The inner product generalizes to:
\begin{equation}
    \langle \Psi | \Phi \rangle = \int dt_1 dt_2 \, \Psi^*(t_1, t_2) \, \Phi(t_1, t_2),
\end{equation}
assigning norm over complexified time. Decoherence emerges as a loss of overlap along $t_2$.

Observation, within this geometry, becomes a slice or hypersurface across $T^2$. Events are not singular spacetime points, but extended structures whose classicality emerges via coherent alignment of internal degrees of freedom. This relational ontology supports the interpretation of time as emergent from subsystem correlations.

Speculatively, one may link the coherence timescale $1/k_2$ to cognitive processes. If awareness requires sustained quantum coherence---as hypothesized in Penrose–Hameroff models---then the $T^2$ geometry may underlie not only physics but consciousness. Temporal experience would reflect branching in $t_2$ modulated by perception across $t_1$.

Causal structure in $T^2$ supports both forward and retrocausal phenomena. Locality in $t_1$ preserves standard relativistic cones, while coherence-mediated correlations in $t_2$ allow for time-symmetric or post-selected effects without violating unitarity.

Ultimately, time in this framework is not an external parameter, but a dynamic, compactified field. Its geometry encodes causality, coherence, entropy, and cognition---a complete, self-consistent structure in which the dual flows of time reconcile quantum and classical worlds.

\section*{Final Remarks}
The two-time manifold $T^2$ establishes a comprehensive foundation for reinterpreting time not as an external parameter but as a compact, quantized, and dynamically structured entity. Within this framework, classical causality and quantum coherence emerge as dual aspects of a unified temporal geometry, embedded in a holomorphic structure that supports both unitary evolution and non-unitary processes such as decoherence and measurement.

By promoting time to a compact complex surface, this work provides a setting where quantum interference, entropy growth, and measurement collapse become manifestations of geometry rather than postulates. The extended Schrödinger and Lindblad equations defined on $T^2$ offer a natural language for incorporating both deterministic evolution and stochastic projection as geometric flows.

Central to the formalism is the para-Hermitian structure on the generalized tangent bundle, which encodes T-duality, modular symmetry, and coherence dynamics in a covariant framework. Its close relationship to generalized complex geometry supports a broad class of dualities--including temporal winding-momentum duality--that underlie black hole entropy, modular Hamiltonians, and holographic entanglement.

Through compactification of the coherence direction $t_2$, we recover a minimal temporal resolution $\Delta t_2 \sim \sqrt{\alpha'}$, introducing a lattice of quantized temporal modes. This structure regulates ultraviolet divergences, modifies the Einstein equations through $t_2$-averaged stress-energy, and induces non-commutative temporal geometry at the Planck scale.

Furthermore, we have shown how temporal entanglement violates generalized Bell inequalities, with interference modulated by spectral phase factors $\exp(i \Delta E t_2 / \hbar)$. The measurement process is reinterpreted as a modular projection onto classical time slices, consistent with open-system decoherence and holographic modular flow.

Experimental implications of this framework span ultrafast optics, neutrino time-of-flight dispersion, and black hole thermodynamics. Signatures include attosecond-scale decoherence jitter, frequency sidebands from winding modes, and shifted thresholds in cosmic ray and gamma-ray spectra. These predictions offer a testable window into the quantum structure of time.

In sum, the $T^2$ framework recasts measurement, entropy, entanglement, and gravitation as emergent from a complexified temporal manifold. It dissolves the dichotomy between coherence and collapse, between causality and probability, and between spacetime and information. Time, in this view, is not an absolute background--but a compact, relational, and dynamical object woven into the quantum and gravitational fabric of the universe.

Future work should refine this formalism in loop quantum gravity, double field theory, and AdS/CFT tensor networks, and investigate new classes of experimental probes sensitive to coherence-scale dynamics. Ultimately, the geometry of time may hold the key to resolving the deepest puzzles of physics--from quantum foundations to the origin of cosmological order.

\bibliographystyle{abbrv}
\bibliography{bibliography}

\end{document}